\documentstyle[aps,prl,psfig,floats]{revtex}
\begin{document}
\draft
\twocolumn[\hsize\textwidth\columnwidth\hsize
\csname @twocolumnfalse\endcsname
\title{Theoretical analysis of
Cooper-pair phase fluctuations in underdoped cuprates:\\
a spin-fluctuation exchange study}
\author{D. Manske$^1$, T. Dahm$^{2}$\cite{pa}, and K.H. Bennemann$^1$}
\address{$^1$Institut f\"ur Theoretische Physik, Freie Universit\"at Berlin,
Arnimallee 14, D-14195 Berlin, Germany}
\address{$^2$Max-Planck-Institut f\"ur Physik komplexer Systeme,
N\"othnitzer Str. 38, D-01187 Dresden, Germany}
\date{\today}
\maketitle
\begin{abstract}
We study Cooper-pair phase fluctuations in cuprate superconductors
for a spin fluctuation pairing interaction.
Using an electronic theory we calculate in particular for the
underdoped cuprate superconductors the superfluid density
$n_s(T)$, the superconducting transition temperature $T_c(x)
\propto n_s$ below which phase coherent Cooper-pairs
occur, and $T_c^{*}(x)$ where the phase incoherent Cooper-pairs
disappear. Also we present results for the penetration depth
$\lambda(x,T)$ and for the weak pseudogap temperature
$T^{*}(x)$ at which a gap structure occurs
in the spectral density. A Mei{\ss}ner
effect is obtained only for $T < T_c$.
We find that phase fluctuations become increasingly important
in the underdoped regime and lead to a reduction of $T_c$ in
good agreement with the experimental situation.
\end{abstract}
]
\narrowtext

In accordance with the experimental observation for underdoped
cuprate superconductors that the superconducting transition
temperature is related to the superfluid density $n_s$ and behaves
as $T_c \propto n_s(0)$, in contrast to the overdoped case
\cite{Uemura}, we
use an electronic theory which takes into account phase
fluctuations of the Cooper-pairs. From this one expects a phase diagram
with  $T_c^{*}$ at which phase incoherent Cooper-pairs occur
and a lower transition temperature
$T_c \propto n_s$ at which these Cooper-pairs become
phase coherent. Such a physical picture for the underdoped
cuprates has been treated phenomenologically before by
Chakraverty {\it et al.} \cite{Chakraverty}, more recently by
Emery and Kivelson \cite{EmeryKivelson}, and also by
Schmalian {\it et al.} \cite{Schmalian}.

Close to $T_c$ one gets as in Ginzburg-Landau (GL) theory the free
energy (without external field)
\begin{equation}
\Delta F = F_S - F_N = \Delta F_{BCS} + \Delta F_{phase}
\quad ,
\label{eq:freeenergy}
\end{equation}
where $\Delta F_{BCS}=\frac{1}{2}N(E_F)\Delta_0^2$ and
$\Delta F_{phase}=\frac{\hbar^2}{2m^{*}}\bar{n}_s$
with $\bar{n}_s=n_s\langle\nabla\Phi({\bf r})\nabla\Phi(0)
\rangle$.
Here, we assume only a spatial dependence of the phase $\Phi({\bf r})$
of the superconducting order parameter. This implies a
superconducting transition at $k_BT_c=\Delta F_{BCS}$, when
$\Delta F_{BCS} < \Delta F_{phase}$ and Coper-pairs are
formed, and $k_BT_c=\Delta F_{phase}
\propto n_s(0)$, when $\Delta F_{phase}<\Delta F_{BCS}$ and
Cooper-pairs become phase coherent.
This neglects 2d effects.
In this letter we use an electronic theory in order to calculate
the doping dependence of $N(E_F)$, $\Delta_0(x)$, and $n_s(x)$
and thus the two contributions to $\Delta F$ and the two doping
regimes for non-BCS and BCS behavior of the superconducting
transition temperature.
Based on a spin fluctuation exchange mechanism we determine the
doping dependence of the phase diagram with $T_c^{*}(x)$, $T_c(x)$,
and also $T^{*}$ at which far above $T_c$ a gap appears in the
spectral density.
In accordance with Eq. (\ref{eq:freeenergy}) we find from our
microscopic calculation that phase fluctuations
of the Cooper-pairs are significant for the underdoped cuprates.
For the overdoped superconductors we obtain a mean-field behavior
and $T_c^{exp} \simeq T_c^{*}$. Our results suggest (see also
\begin{figure}[t]
\vspace{-1.0cm}
\centerline{\psfig{file=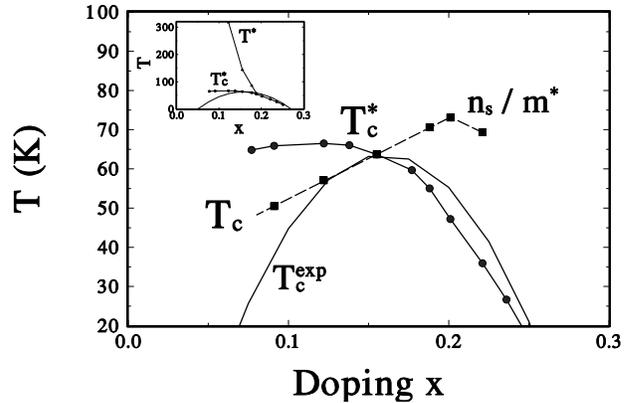,width=8.7cm,angle=-90}}
\vspace{1ex}
\caption{Phase diagram for high-$T_c$ superconductors
resulting from a spin fluctuation induced Cooper-pairing
including their phase fluctuations.
$T_c^{*}$ denotes the temperature below which Cooper-pairs are formed.
The dashed curve gives $T_c \propto n_s(T=0,x)$.
Below $T^{*}$ (see inset) we get a gap structure in the spectral
density.}
\label{fig1}
\end{figure}
Kosterlitz-Thouless (KT) theory) that a Mei{\ss}ner
effect occurs only for $T < T_c(x)$, since $\lambda^{-2}
\propto \bar{n}_s$.

 We treat the cuprates by using as a model the 2D Hubbard
Hamiltonian for a $\mbox{CuO}_2$-plane
\begin{equation}
H = - \sum_{\langle ij \rangle \, \sigma}
t_{ij}\left( c_{i\sigma}^+ c_{j\sigma} +
c_{j\sigma}^+ c_{i\sigma}\right)
+ U\, \sum_i n_{i\uparrow}n_{i\downarrow}
\quad .
\label{eq:hubbard}
\end{equation}
Here, $c_{i\sigma}^+$ creates an electron with spin $\sigma$
on site $i$, $U$ denotes the on-site Coulomb interaction,
and $t_{ij}$ is the hopping integral.
From $H$ and the functional differentiation of the free
energy $F$ with respect to the Green's function
${\cal G}$, $\delta F\{H\} / \delta {\cal G} = \Sigma$,
one obtains with the help of the Dyson equation the self-energies.
Note, this gives in general coupled equations for the amplitude and
phase $\Phi({\bf r})$ of ${\cal G}$ \cite{Schaefer}
and in accordance with the GL (Wannier) type
treatment an energy gain due to phase coherence of Cooper-pairs
and due to Cooper-pair condensation \cite{Chakraverty,EmeryKivelson,%
Schaefer,Remark1}.

 It follows from Eq. (\ref{eq:hubbard}) that the
quasiparticle self-energy components $X_{\nu}$ ($\nu=
0$, $3$, $1$) with respect to the Pauli matrices $\tau_{\nu}$
in the Nambu representation \cite{Nambu}
are given by
\begin{eqnarray}
X_{\nu}({\bf k},\omega)
& = &
N^{-1}\sum_{{\bf k'}}\int_0^{\infty}
d\Omega \, \left[ P_s({\bf k} - {\bf k'},\Omega) \pm
P_c({\bf k} - {\bf k'},\Omega)\right]
\nonumber\\
& \times &
\int_{-\infty}^{\infty} d\omega' \, I(\omega,\Omega,\omega')
A_{\nu}({\bf k'},\omega')
\quad .
\label{eq:selfenergy}
\end{eqnarray}
 Here, we used the $T$-matrix
\cite{Tewordt} or fluctuation exchange (FLEX) approximation
\cite{Bickers,Pao,Monthoux,DahmTewordt} in which a
Berk-Schrieffer-like \cite{BerkSchrieffer} pairing interaction
is constructed with the dressed one-electron Green's functions.
The spin fluctuation interaction is given by
$P_s = \left( 2\pi\right)^{-1}$ $U^2 \,\mbox{Im }
\left(3\chi_s - \chi_{s0}\right)$ with
$\chi_s = \chi_{s0}\left( 1 - U\chi_{s0}\right)^{-1}$
and the charge fluctuation interaction is
$P_c = \left( 2\pi\right)^{-1}U^2\,\mbox{Im }
\left(3\chi_c - \chi_{c0}\right)$ with
$\chi_c = \chi_{c0}\left( 1 + U\chi_{c0}\right)^{-1}$,
where $\mbox{Im }\chi_{s0,c0}({\bf q},\omega)$ is given in
Ref. \onlinecite{DahmTewordt}.
The subtracted terms in $P_s$ and $P_c$
remove a double counting that occurs in second order.
In Eq. (\ref{eq:selfenergy}) the plus sign holds for
$X_0$ (quasiparticle renormalization) and $X_3$ (energy shift),
and the minus sign for $X_1$ (gap parameter).
The kernel $I$ and the spectral functions $A_{\nu}$
are given
in Ref. \onlinecite{DahmTewordt}.
%

For the numerical evaluation we use a
bare tight-binding dispersion relation
$\epsilon({\bf k})=2t\left[ 2 - \right.$ $\left.\cos(\pi k_x) -
\cos(\pi k_y) - \mu \right]$ and $U/t=4$. Then, the doping dependence
$n=\frac{1}{N}\sum_{\bf k}n_{\bf k}=1-x$ is
determined with the help of the ${\bf k}$-dependent occupation
number $n_{\bf k}=2\int_{-\infty}^{\infty}d\omega f(\omega)
N({\bf k},\omega)$ that is calculated self-consistently,
where $N({\bf k},\omega)=A_0({\bf k},\omega) + A_3({\bf k},\omega)$.
$n=1$ corresponds to half filling.
Our numerical calculations are performed on a square lattice
with $256$x$256$ points in ${\bf k}$ space of the Brillouin zone
and with $200$ points on the real $\omega$ axis up to $16t$ with
an almost logarithmic mesh. The full momentum and frequency dependence
of the quantities is kept \cite{SereneHess}.
Thus, our calculation includes pair breaking effects on the
Cooper-pairs resulting from lifetime effects of the elementary
excitations.
$T_c^{*}$ is determined from the linearized gap equation and the
superconducting state is found to have $d$-wave symmetry
\cite{DahmTewordt}.
The transition temperature $T_c$ at which phase coherence occurs
is determined by $T_c \propto n_s(x)$
where the superfluid density $n_s(x,T)/m$ is calculated
self-consistently from
\begin{displaymath}
\frac{n_{s}}{m} = \frac{2t}{\hbar^2}\, \left(S_N - S_S\right)
\quad \mbox{with} \quad
S_N = \frac{\hbar^2c}{2\pi e^2t}\int_0^{\infty}\sigma_1(\omega)\,
d\omega .
\end{displaymath}
$S_S$ is the corresponding expression in the superconducting state.
Here, we utilize the f-sum rule for
the real part of the conductivity $\sigma_1(\omega)$, i.e.,
$\int_0^{\infty}\sigma_1(\omega)\, d\omega =
\pi e^2n / 2m$ where $n$ is the 3D electron density and $m$
denotes the effective band mass for the tight-binding band
considered. $\sigma(\omega)$ is calculated in the normal
and superconducting state using the standard
Kubo formula \cite {Wermbter}.
Vertex corrections have been neglected.
Physically speaking, we are looking for the
loss of spectral weight of the Drude peak at $\omega=0$ that
corresponds to excited quasiparticles above the superconducting
condensate for temperatures $T < T_c^{*}$.
Finally, the penetration depth $\lambda(x,T)$ is calculated
within the London theory through $n_s \propto \lambda^{-2}$
\cite{London}.

In Fig. \ref{fig1} we show the phase diagram for the cuprate
superconductors obtained from our electronic theory and compare
with the generalized experimental phase diagram $T_c(x)$ that
describes many superconductors as pointed out by Tallon \cite{Tallon}.
We get $T_c \approx T_c^{*}$ for
$x > x_{opt}$, whereas for the underdoped
superconductors $T_c \propto n_s(T=0)$
agrees much better with the experimental results than $T_c^{*}$.
Note, for the overdoped cuprates one gets the superconducting
transition at $T_c^{*}$, where phase coherent Cooper-pairing
occurs since
$\Delta F_{phase}$ becomes largest.
$T_c^{*}$ results from $\phi(x,T_c^{*})=0$, where $\phi(k,\omega)=
Z(k,\omega)\Delta(k,\omega)$ is the strong-coupling superconducting
order parameter and is a mean-field
result in the sense that Cooper-pair phase
fluctuations have been neglected. However, the
\begin{figure}[t]
\vspace{-1.0cm}
\centerline{\psfig{file=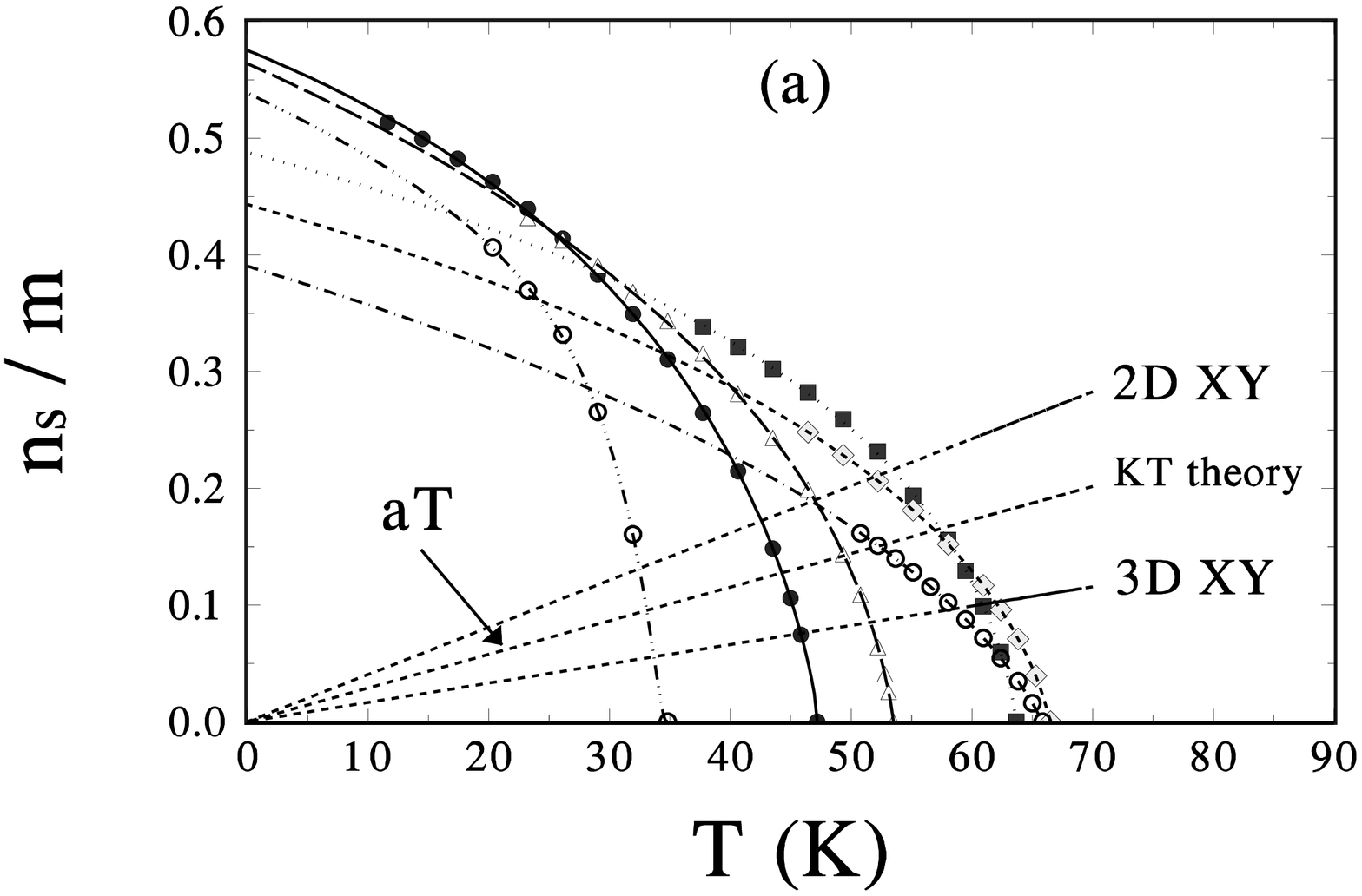,width=8.7cm,angle=-90}}
\vspace{-1.0cm}
\centerline{\psfig{file=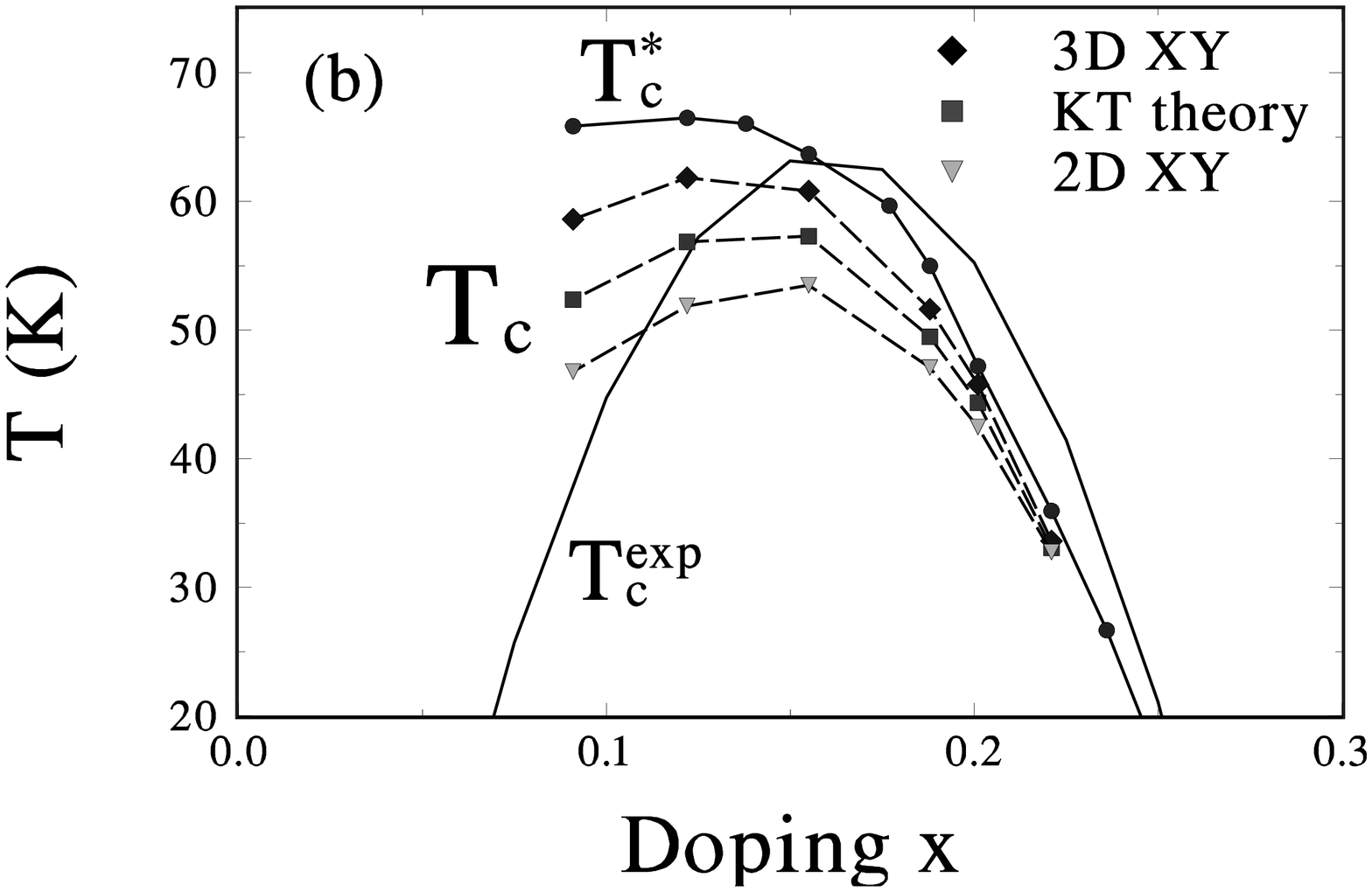,width=8.7cm,angle=-90}}
\vspace{1ex}
\caption{(a) Results for the superfluid density divided by $m$
in units $10^{-2}m_e^{-1}\AA^{-2}$ as a
function of temperature $T$ for various doping concentrations:
x=0.221 (open circles), x=0.201 (circles), x=0.188 (triangles),
x=0.155 (squares), x=0.122 (diamonds), x=0.091 (circles).
$n_s$ vanishes at $T=T_c^{*}$.
From $n_s(x,T)$, $T_c$ is given as indicated by the
intersection with $aT$, where $1/a=2.202$ (3D XY), $1/a=\pi/2$ (KT
theory), $1/a=0.9$ (2D XY). The obtained results for $T_c$ are
shown in (b).}
\label{fig2}
\end{figure}
AF fluctuations are treated well beyond the
mean-field approximation. It is
remarkable that we find approximately that $T_c$ varies linearly
with $x$. This cannot be the case for $n_s(x \to 0)$.
We also expect that an improved inclusion of the pseudogap behavior
and reduction of the density of states (DOS) due to
antiferromagnetic (AF) correlations will yield a faster decrease
of $T_c(x)$ for $x \to 0$. Also important is the fact that we find
an optimal $T_c$ for $x \simeq 0.15$ since for $x < x_{opt}$ we find
$T_c \propto n_s(T=0)$ and for $x > x_{opt}$ one has
$T_c \propto \Delta(T=0)$. Of interest are also the
results for the pseudogap temperature $T^{*}$ at which a gap structure
in the DOS in the normal state occurs. This has been
discussed in Refs. \onlinecite{DahmTewordt,Schmalian}. Such a
behavior is seen experimentally in various SIN tunneling data,
reflectivity measurements as well as in the two-magnon response
in Raman scattering \cite{Timusk,Ruebhausen}.
Physically speaking, below $T^{*}$ the electrons
are strongly coupled to the magnetic degrees of freedom which are
generated due to paramagnon (spin fluctuation) excitations in the
system. Finally, we would like to mention that at
$x^*=0.19$, $T^*$ and $T_c^*$ coincide.
Furthermore, our results indicate that for $x > x^*$ the non-BCS
behavior $T_c \propto n_s(T=0)$ seems not valid anymore.
In contrast to the underdoped regime, there
the energy gain due to Cooper-pair formation is 
smaller than the energy to break up phase coherence.
Thus, Cooper-pair phase fluctuations are unimportant in this regime.
Note, $T_c$ may be viewed in analogy to the Curie temperature
in Ferromagnets like Ni or Fe, where spin-disorder occurs.

In order to investigate  different models used in the context of
Cooper-pair phase fluctuations
\cite{Chakraverty,EmeryKivelson,Schmalian}, we discuss now the
superfluid density for finite temperatures.
In Fig. \ref{fig2}(a) results are given for $n_s(T,x)/m$.
To obtain $n_s(T=0)$ we have used a polynomial fit up
to third order \cite{Remark2}.
For the doping values investigated here, we find a linear behavior
of $n_s(0,x)$ except for the strongly
overdoped case, i.e. $x \geq 0.22$ where $n_s(T=0)$ starts to
decrease. This behavior indicates a crossover to the
(BCS) weak-coupling limit where $n_s=n$ for $T=0$ is expected.
This implies that $\lambda(T=0,x)$ is asymmetric with respect
to optimal doping as suggested from Fig. \ref{fig1}.
Note that the energy $\Delta F_{phase} \propto n_s(0)$ is still
larger than the energy gain due to Cooper-pair condensation.
Thus we conclude that Cooper-pair phase fluctuations are
unimportant in the overdoped regime.
Note, $n_s(T,x) \to 0$ for $T \to T_c^{*}$, since Cooper-pairs
disappear at $T_c^{*}$.
However, the phase coherence temperature $T_c$ has to be determined
by spatially averaging over the Cooper-pair phase fluctuations.
In the presence of spatial phase fluctuations
the average superfluid density $\bar{n}_s$ will vanish at $T_c$ so
that as in KT theory no Mei{\ss}ner effect occurs above $T_c$.
Note, within KT theory $T_c$ is given by
\begin{equation}
\frac{1}{a} \, \hbar^2 \, \frac{n_s(T_c)}{4m}
= k_B T_c
\quad ,
\end{equation}
where $1/a=\pi/2$. In the case of the similar 2D $XY$ model
one has $1/a=0.9$. Our construction used in Fig. \ref{fig2}
involves only the
determination of $n_s(T=T_c)$, but does {\it not} imply a jump in $n_s$
at $T_c$ which is true only for the 2D $XY$-model and KT theory. For a
comparison with Ref. \onlinecite{Schmalian},
we show also results for the 3D version of the $XY$-model.
All these theories take into account phase and amplitude of the
Cooper-pairs. We see that in the 2D (3D) version of the $XY$-model
$T_c$ is lower (higher) compared to the KT theory.
Physically speaking, in the $XY$-model $n_s$ (or the phase stiffness)
is the only relevant energy scale. Thus one always has $T_c
\propto n_s$. In the KT theory one has two energy scales, namely
the phase stiffness and also the vortex core energy.
Nevertheless, all these theories yield within our
microscopic treatment an optimal doping concentration.
Note, we obtain $n_s(x \to 0)$ due to pair breaking resulting
from the decreasing lifetimes of the Cooper-pairs \cite{RemarkDensity}.
Improved calculations taking into account AF correlations leading
to static antiferromagnetism cause a further decrease of $n_s(x)$
for $x \to 0$ \cite{Manske}.

\begin{figure}[t]
\vspace{-1.0cm}
\centerline{\psfig{file=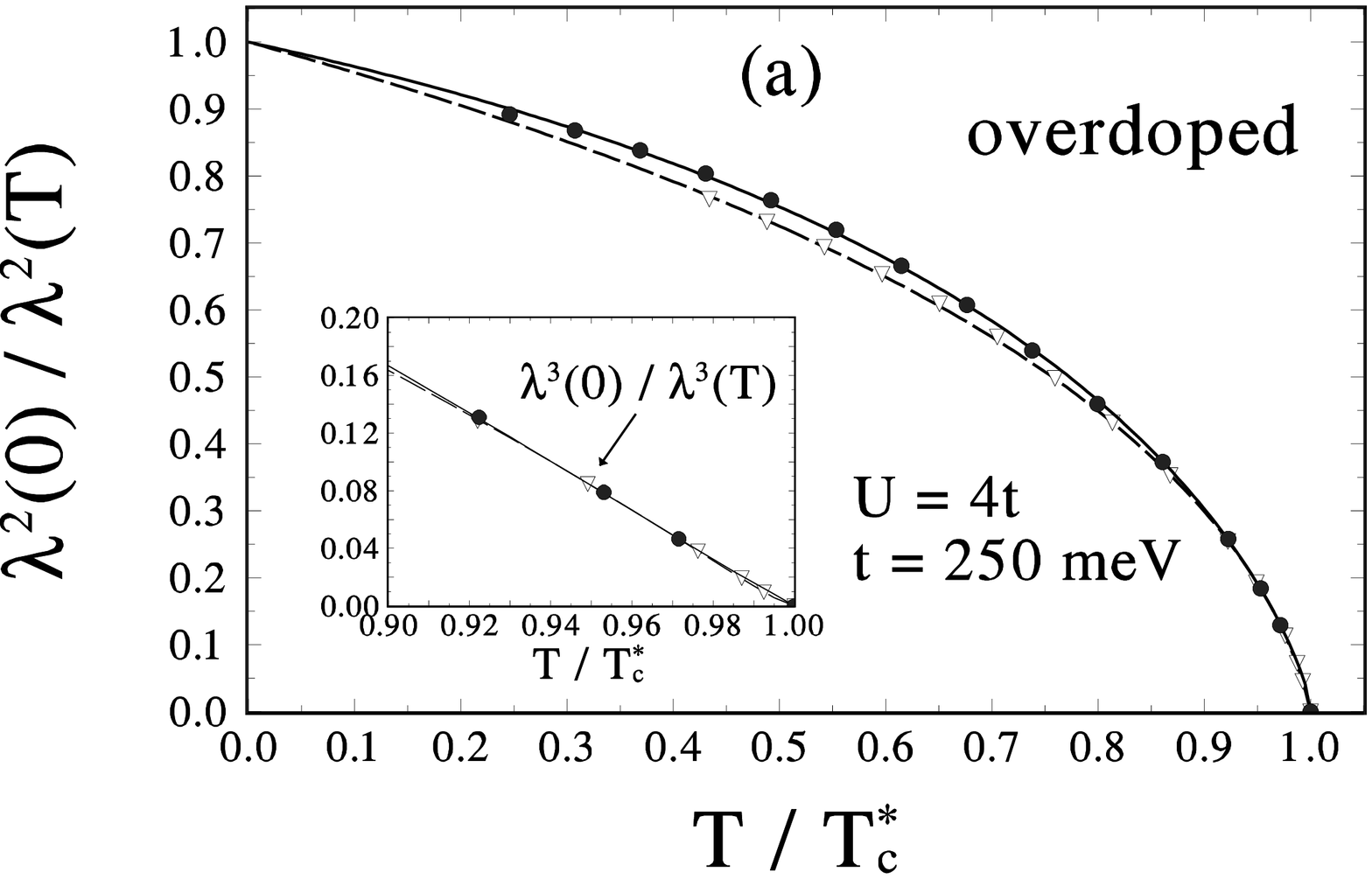,width=8.7cm,angle=-90}}
\vspace{-1.0cm}
\centerline{\psfig{file=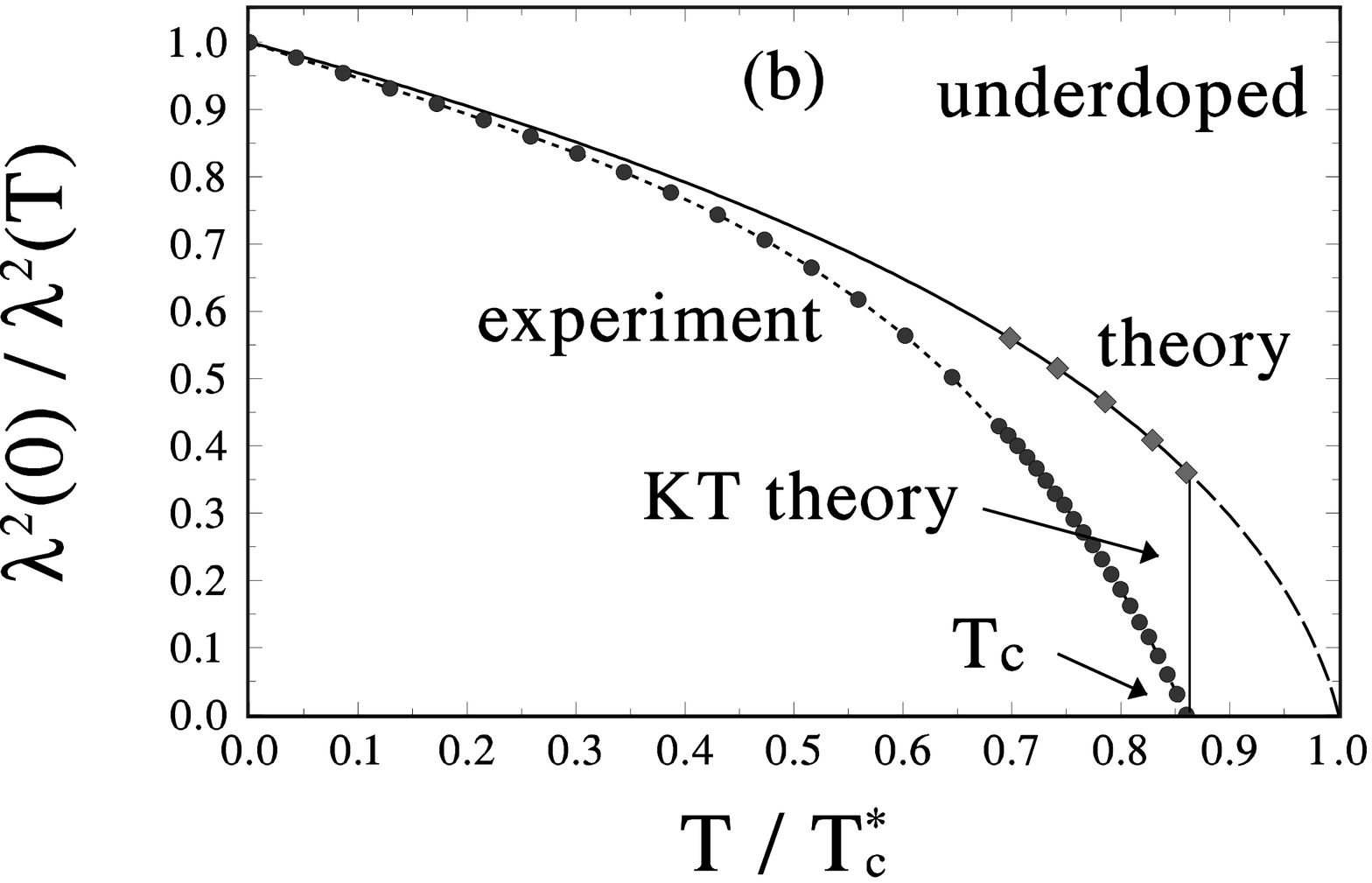,width=8.7cm,angle=-90}}
\vspace{1ex}
\caption{Results for the penetration depth $\lambda(T,x)$
($n_s \propto \lambda^{-2}$). The inset in (a) shows a remarkable
linear behavior of $\lambda^{-3}(T)$ for $T \simeq T_c^{*}$ even
without having included critical fluctuations.
In (b) we indicate the behavior expected for $\bar{n}_s$ and
derived from KT theory, where $\bar{n}_s(T) \to 0$ at $T_c$.}
\label{fig3}
\end{figure}

%
In Fig. \ref{fig3} we show the mean field results for the
penetration depth.
In (a) the overdoped case is displayed: our results for
$\lambda^{-2}(T)$ agree qualitatively with the experimental data
\cite{Kamal,Panagopoulos}.
It is remarkable that we find a linear behavior
of $\lambda^{-3}(T)$ (see inset).
In Ref. \onlinecite{Kamal} at $T=0.9T_c^{*}$ a value for
$\lambda^3(0)/\lambda^3(T)$ of $0.2$ has been found. Instead, we
find 0.16. Note that we did not include critical
fluctuations. A closer inspection of our data
leads to the conclusion that the rapid opening of
the superconducting gap -which is calculated self-consistently-
is the main reason for the reported behavior of the
penetration depth \cite{Pao,Dahm}.
Thus, we conclude that the existence of
critical fluctuations is not necessary in
order to understand the observed behavior of the penetration depth
close to the critical temperature in the cuprates.
This might shed some light on the significance of
vertex corrections.
However, very close to $T_c$ we cannot present numerical results
yet. For this range it remains unclear whether or not a change
of the powerlaws from $1/3$ to $1/2$ occur.
Fig. \ref{fig3}(b) indicates that
for calculating $\lambda(T,x)$ one must use the
superfluid density $\bar{n}_s$
referring to phase coherence as is done in KT theory. Note, static
KT theory predicts an universal jump in $\bar{n}_s$ at $T=T_c$ as
indicated, which is not observed experimentally
\cite{Kamal,Panagopoulos,Corson};
the $n_s$ results obtained from our electronic theory refer only
to the Cooper-pair density $n_s(x,T)$ but $\bar{n}_s = n_s$
at $T < T_c$. As mentioned earlier, the spatially phase averaged
$\bar{n}_s(x,T)$ yields $\bar{n}_s(T_c,x) = 0$ for incoherent
Cooper-pairs. The comparison with experiment indicates that $\bar{n}_s
< n_s$ already for $T < T_c$. No Mei{\ss}ner effect is expected for
$ T_c < T < T_c^{*}$.
While our calculation of $n_s(T,x)$ is expected to give the
correct tendency for the behavior of the underdoped superconductors,
stronger AF effects must be included for $x \to 0$. Also,
quantum phase fluctuations are neglected.
All this would cause a more rapid decrease
of $T_c \propto n_s(x)$ in Fig.\ref{fig1}.

In summary, it is interesting that our electronic theory using
the model Hamiltonian $H$,
Eq. (\ref{eq:hubbard}), gives remarkable agreement with some
basic experimental observations.
In particular, we have shown within a {\it microscopic} theory
based on a spin fluctuation pairing mechanism
that Cooper-pair phase fluctuations become important for underdoped
cuprates.
This gives a microscopic justification for the phenomenological
theories in Refs. \onlinecite{Chakraverty,EmeryKivelson}.
Of course, a fully self-consistent
determination including the phase fluctuations of the Green's
functions and the electronic exitations should be
performed \cite{Manske}.
Thus, one might get also an insight about the origin of the
pseudogaps above $T_c$ and revealing asymmetric doping dependence
of various properties with respect to $x_{opt}$.

We acknowledge helpful discussions with F. Sch\"afer and C. Timm.


\begin{references}
\bibitem{Remark2}It might be desirable to calculate more values of
$n_s(x)$ as $x \to 0$. However, the self-consistent solution
of Eq. (\protect\ref{eq:selfenergy}) becomes more
difficult for $T \to 0$.
\bibitem{RemarkDensity}
It is important that we find for the underdoped cuprates
approximately $n_s/n \simeq 0.3$ for $T \to 0$. Then
the Cooper-pair wave function has a radius that is
comparable or smaller than the
distance between Cooper-pairs. This supports
our physical picture, $T_c \propto n_s$, and the significance
of Cooper-pair phase fluctuations.
\end{references}
\end{document}